\documentclass[runningheads,fleqn]{cl2emult}

\usepackage{makeidx}  
\usepackage{graphicx} 
\usepackage{subeqnar} 
\usepackage{multicol} 
\usepackage{cropmark} 
\usepackage{phys}     
\makeindex            

\newcommand{\cL}{{\mathcal L}}

\begin{document}
\title*{Beyond Wavelets: \protect\newline Exactness theorems for physical calculations}
\toctitle{Beyond Wavelets: \protect\newline Exactness theorems for physical calculations}
%
%
\titlerunning{Beyond Wavelets}
%
\author{T.A. Arias
\and T.D. Engeness}
\authorrunning{T.A. Arias and T.D. Engeness}
%
%
\institute{Department of Physics\\ Massachusetts Institute of
Technology, Cambridge MA 02139, USA}

\maketitle              

\begin{abstract}
This paper develops the use of wavelets as a basis set for the
solution of physical problems exhibiting behavior over wide-ranges in
length scale.  In a simple diagrammatic language, this article reviews
both the mathematical underpinnings of wavelet theory and the
algorithms behind the fast wavelet transform.  This article
underscores the fact that traditional wavelet bases are fundamentally
ill-suited for physical calculations and shows how to go beyond these
limitations by the introduction of the new concept of semicardinality,
which allows basic physical couplings to be computed {\em exactly}
from very sparse information, thereby overcoming the limitations of
traditional wavelet bases in the treatment of physical problems.  The
paper then explores the convergence rate of conjugate gradient
solution of the Poisson equation in both semicardinal and lifted
wavelet bases and shows the first solution of the Kohn-Sham equations
using a novel variational principle.

\end{abstract}

\section{Introduction}

Problems in the physical sciences often involve behavior spanning many
length scales.  Three ingredients, which do not necessarily follow one
from the other, are needed to deal effectively with such problems :
(1) compact representation of the fluctuations of physical fields over
different length scales in different regions of space, (2) economical
expression of the physical couplings among such fields, and (3)
efficient means for solving the resulting equations.  The discussion
below gives a very powerful general means for dealing with such
problems.  Although, the discussion is general, for concreteness we
shall consider the calculation of electronic structure of matter,
reviewed briefly in Sec.~\ref{sec:elec}, as a prototypical example.

As we shall see in Sec.~\ref{sec:MRA}, the central concept underlying
wavelet theory, multiresolution analysis, is a very elegant and
powerful mathematical tool for providing compact representations, but
the expression of the most common physical couplings is very awkward
in traditional wavelet bases.  Sec.~\ref{sec:SemiC} shows how the
introduction of a new concept, {\em semicardinality} provides a means
of overcoming this limitation and providing an extremely efficient
and, surprisingly, {\em exact} means of expressing the two most
fundamental physical couplings.  Finally, Sec.~\ref{sec:solutions}
shows extremely efficient methods for solving Poisson's equation and
the Kohn-Sham equations within this new framework.

\section{Electronic structure of matter} \label{sec:elec} 

\subsection{Multiscale nature}

\begin{figure}
\includegraphics[width=0.75\textwidth]{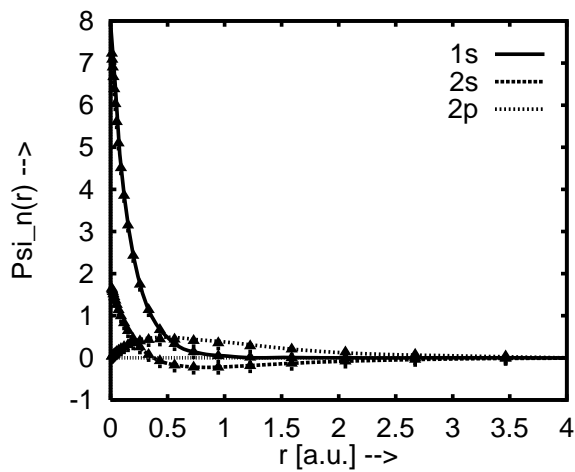}
\caption{Kohn-Sham orbitals of the carbon atom within the local
density approximation from standard atomic software ({\it symbols\,})
and multiresolution analysis({\it curves\,})~\protect\cite{mgras} }
\label{fig:C}
\end{figure}

It is well known that the electronic wave functions in molecular and
condensed-matter systems vary much more rapidly near the atomic nuclei
than in interatomic regions.  In the immediate vicinity of the nucleus
and its strong attractive potential, the electrons possess large
kinetic energies, as reflected by high spatial frequencies evident in
the orbitals.  Figure~\ref{fig:C} illustrates this behavior, using the
carbon atom as an example.  The curves in the figure show the
Kohn-Sham orbitals of the atom as computed within the local density
approximation \cite{KohnSham} to density functional theory
\cite{HohenbergKohn}.  The high-frequency ``core'' region extends only
approximately 0.5~bohr radii out from the nucleus, beyond which the
variations in the wave functions are quite smooth.  Resolving the
cusps in the $s$ states of this atom requires a resolution on the
order of 0.03~bohr (corresponding to a plane wave cutoff \cite{bible}
of nearly 160,000~rydberg).  To provide this resolution uniformly
throughout the computational cell of these calculations, which is
8~bohr on a side, would require a basis with 16~million coefficients.
The vast majority of these basis functions would be wasted as they
would serve to provide unnecessarily high resolution outside the core
region.

The issue of multiple length-scales in electronic structure is not
new.  It has driven the development of a variety of techniques which
are now quite mature, including the atomic sphere family of
approaches, which uses one type of basis set inside of a set of
spheres organized around the nuclei and another type of basis set
outside of the spheres, and the plane wave pseudopotential approach,
which replaces the atomic core with an effective potential
manufactured to have similar scattering properties.  While each of
these approaches has had great success, none is systematically
improvable to complete convergence in a simple, practical manner, and
each requires great care and expertise in the selection and
construction of the atomic spheres or in the development of
appropriate pseudopotentials.  As a result, a general method is still
needed to obtain unambiguous results of sufficient accuracy to permit
direct and systematic study of the relative accuracy of competing
density functionals and alternate theories of electronic structure.
This situation calls for precisely the capabilities of multiresolution
analysis, which holds the promise of at last enabling the systematic
evaluation of different theories of electronic structure at high
precision.

\subsection{Electronic Structure} \label{sec:Lagr}

Throughout this paper, we shall work within the local density
approximation (LDA) to density functional theory\cite{KohnSham}.  A
novel way to express the Kohn-Sham equations within this
approximation, first introduced in \cite{JCP}, is through the
following variational principle (in atomic units
$\hbar=m=e=1$),
\begin{eqnarray}
0 = \delta {\mathcal L}_{LDA} & = & \delta \int \left[ \sum_i f_i
\frac{1}{2} {|\nabla \psi_i(\vec{r})|^2} + V_{\mbox{nuc}}(\vec{r})
n(\vec{r}) \right. \label{eqn:saddle} \\
& & \left. + \, \epsilon^{\mbox{xc}}(n(\vec{r})) n(\vec{r}) + \phi(\vec{r}) n(\vec{r})
- \frac{1}{8\pi} |\nabla \phi(\vec{r})|^2 \right]
d^3\vec{r}. \nonumber
\end{eqnarray}
Here, the minimization is over orthonormal sets of Kohn-Sham orbitals
$\{\psi_i(\vec{r})\}$ and the electrostatic
(Hartree) potential $\phi(r)$ of the electrons, 
$V_{\mbox{nuc}}(\vec{r})$ is the potential of the nuclei, the
electron density is defined as
\begin{eqnarray}
n(r) & \equiv & \sum_i f_i |\psi_i(r)|^2,  \label{eqn:defnsaddle}
\end{eqnarray}
the $f_i$ are the occupancies of the Kohn-Sham orbitals, and
$\epsilon^{\mbox{xc}}(n)$ is the exchange-correlation energy per
electron in a uniform electron gas of density $n$, a highly non-linear
function.  The advantage of this new variational principle is that it
does not involve the traditional long-range Hartree integral,
\begin{eqnarray}
E_{\mbox{Hartree}} & = & \int d^3\vec{r} d^3\vec{r}' \, \frac{n(\vec{r})
n(\vec{r}')}{|\vec{r}-\vec{r'}|}.
\end{eqnarray}
It thus expresses the physics of electronic structure in terms of two
simple types of coupling: (1) purely local, but possibly non-linear
couplings, and (2) semi-local quadratic couplings among the fields
through differential operators.  Any successful approach to electronic
structure and other physical problems must specify how to deal
effectively with these two forms of coupling.  This new variational
principle also leads to a novel solution of the Kohn-Sham equations
(Sec.~\ref{sec:Lagrsolve}).

\section{Compact representation of fields} \label{sec:MRA}

\subsection{Two-scale decomposition}

\begin{figure}
(a) \parbox{0.55\textwidth}{\includegraphics[width=0.52\textwidth]{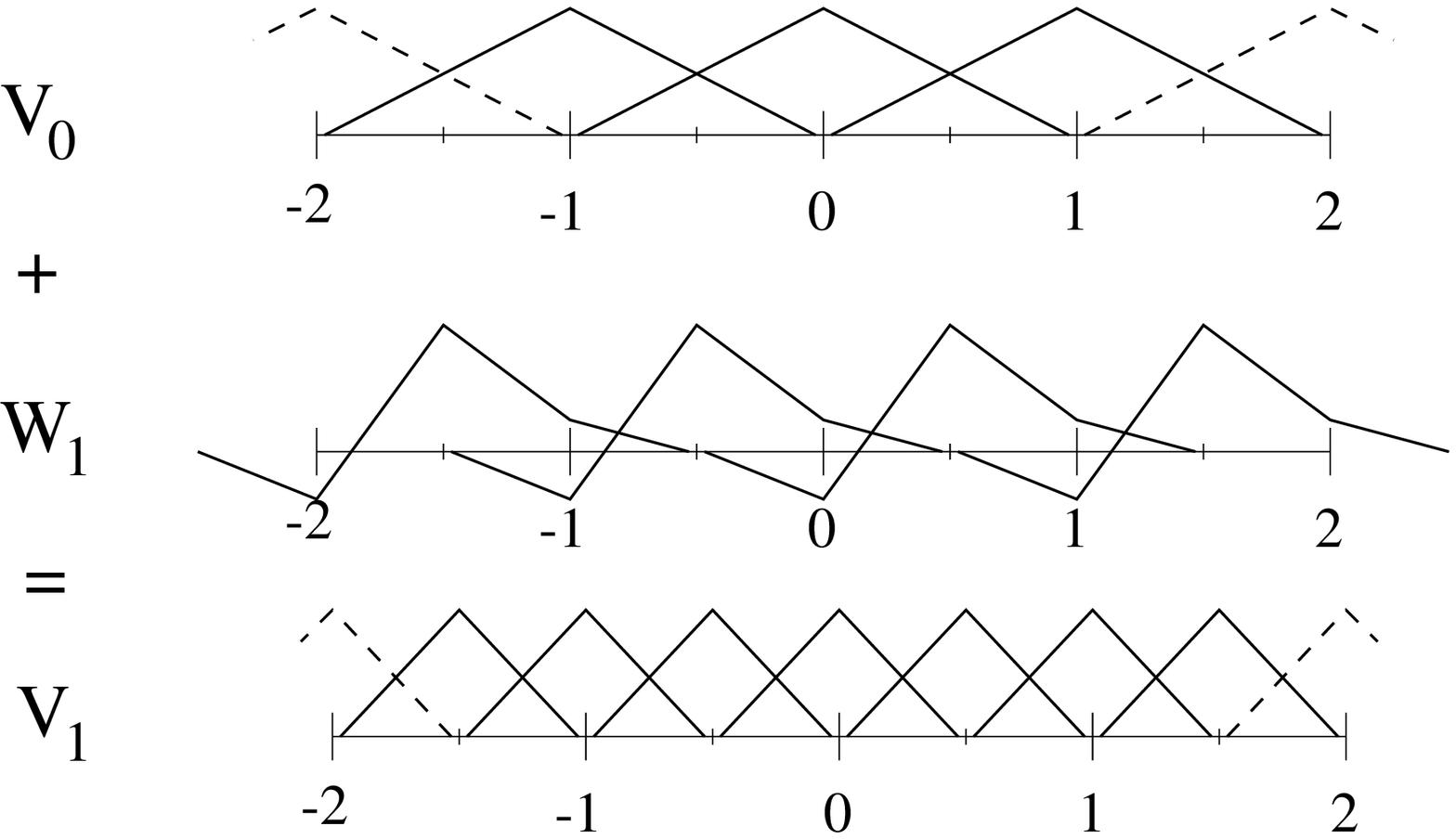}}
(b) \parbox{0.35\textwidth}{\includegraphics[width=0.32\textwidth]{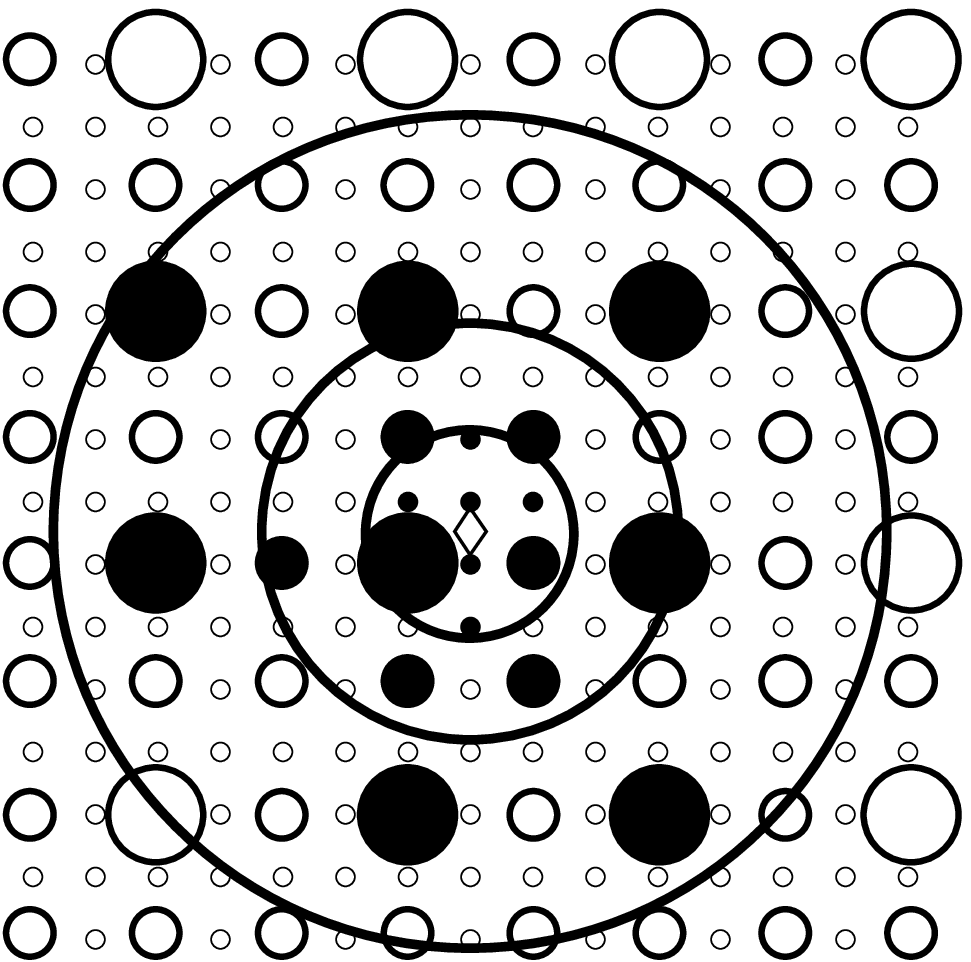}}
\caption{Two-scale decomposition.  ({\bf a}) One dimension. ({\bf b})
Multiple dimensions.  Coarse grid ({\it
larger circles\,}), detail points of finer grids ({\it smaller
circles\,}), atomic nucleus ({\it diamond\,}), spheres of resolution
({\it large circles\,}), surviving basis functions ({\it
filled circles\,})}
\label{fig:MRA}
\end{figure}

Figure~\ref{fig:MRA}a illustrates the concept of {\em two-scale
decomposition}\index{Multiresolution analysis!two-scale decomposition}
for functions in one dimension.  The first row shows a basis set
capable of describing functions which fluctuate on length scales
greater than the unit distance.  This basis consists of copies of a
single function $s(x)$, illustrated as piece-wise linear functions of
triangular appearance in the figure, translated onto each integer
point $n$ along the real axis.  The linear space of functions which
may be written in terms of this basis is called $V_0$.  The simplest
way to double the resolution of this basis is to simply ``compress''
it by a factor of two, reducing the horizontal scale of both the
functions and of the lattice of points on which they sit, as sketched
in the final row of the figure.  The space spanned by this finer basis
is called $V_1$.  Two-scale decomposition builds upon the original
coarser basis for $V_0$ to produce a basis equivalent to the
finer basis for $V_1$ by adding a new set of basis functions $d(x)$
called {\em detail functions}\index{Multiresolution analysis!detail
functions}, or sometimes {\em wavelets}\index{Multiresolution
analysis!wavelets}.  The figure shows these functions along the center
row.  The {\em detail space}\index{Multiresolution analysis!detail
space}, denoted $W_1$, is the space of functions which may be written
solely in terms of these detail functions.  For the combined basis to
be equivalent to the ``compressed'' basis, every function in $V_1$
must be expressible as a sum of one function in $V_0$ and one in
$W_1$.  In the language of linear subspaces, $V_0\oplus W_1=V_1$.

This condition determines much about the form of the basis functions.
First, the spaces $V_0\oplus W_1$ and $V_1$ must have the same
dimension, and thus equal numbers of basis functions, so that there
must be one detail function for each point which appears in the finer
lattice but not in the original coarse lattice.  (In the figure, these
are the odd half-integer points.)  Next the basis functions for both
$V_0$ and $W_1$ must also be in the space $V_1$.  This specifies a
profound restriction on the original basis function, that it may be
written exactly in terms of compressed and translated versions of
itself,
\begin{eqnarray} \label{eqn:tsr}
s(x) & = & \sum_n c_n s(2x-n).
\end{eqnarray}
This relation which connects the same basis functions on two different
scales is known as the {\em two-scale relation}\index{Multiresolution
analysis!two-scale relation}, the coefficients $c_n$ are known as the
{\em two-scale coefficients}\index{Multiresolution analysis!two-scale
coefficients}, and the functions satisfying this condition are known
as {\em scaling functions}\index{Multiresolution analysis!scaling
functions}.  (For a detailed discussion, see\cite{newrmp}.)  Finally,
$V_0\oplus W_1=V_1$ implies that the detail functions are simply
linear combinations of the compressed basis functions,
\begin{eqnarray} \label{eqn:df}
d(x) & = & \sum_n d_n s(2x-n),
\end{eqnarray}
where the coefficients $d_n$ are known as the {\em detail
coefficients}\index{Multiresolution analysis!detail coefficients}.

The preceding conditions ensure that $V_0\oplus W_1 \subset V_1$.  To
show that the combined and the finer spaces are indeed equal, one must
also show that all functions in $V_1$ may be written in terms of a
linear combination of coarse scaling functions and detail functions.
In as much as the two bases contain equal numbers of functions and all
functions in either space are linear combinations of finer scaling
functions, this amounts to showing that a particular linear system of
equations involving the two-scale and detail coefficients always has a
solution.  There is therefore a certain determinant which must be
verified to have non-zero value \cite[Theorem~5.16]{chui}, a condition
easily satisfied in practice.

\subsection{Multiresolution analysis}

To go beyond two-scale decomposition\index{Multiresolution analysis},
we note that starting from $V_0$, one may produce an entire class of
bases of successively higher resolution by reducing the scale of space
by successive powers of two.  This process defines a sequence of
lattices $C_0, \ldots C_N, \ldots$, associated with a sequence of
bases of increasing resolution, $V_0, \ldots V_N, \ldots$.  Simply
changing the horizontal scale in Figure~\ref{fig:MRA}a and the
variable $x$ in Eqs.~(\ref{eqn:tsr},\ref{eqn:df}) by a factor of $2^Q$
gives the prescription for defining the detail space $W_{Q+1}$ which
gives $V_Q + W_{Q+1} = V_{Q+1}$: the basis for $W_{Q+1}$ simply
consists of detail functions compressed by a factor of $2^Q$, $d(2^Q
x)$, centered on the points $D_{Q+1} \equiv C_{Q+1} - C_Q$ which
appear in $C_{Q+1}$ but not $C_Q$.  Finally, iterative application of
two-scale decomposition can produce spaces $V_N$ of {\em arbitrarily}
fine high resolution, $V_N = V_0 \oplus W_1 \ldots \oplus W_{N}$,
and thus express the entire Hilbert space of functions in one
dimension in terms of a basis of hierarchical levels of detail.

These ideas apply equally well in multiple dimensions, as
Figure~\ref{fig:MRA}b illustrates.  Again, the multiresolution
analysis begins with a coarse lattice $C_0$ on which are centered the
coarse scaling functions (large circles in the figure).  Then, to
double the resolution, detail functions appear on the points of $D_1$
(intermediate circles in the figure), and to double the resolution yet
again, detail functions compressed by a factor of two are added to the
points of $D_2$ (small circles in the figure).

Although such a basis separates functions into contributions on
different scales, the final basis still consists of one basis function
for each point on the finest scale grid, thus requiring many
coefficients to represent physical fields.  Figure~\ref{fig:MRA}b
illustrates how to reduce the size of the basis without significant
loss of information.  Only the immediate vicinity of the nuclei
requires the very highest level of resolution $N$, and successively
lower levels of resolution are required as we move outward from the
nuclei.  Therefore, about each nucleus we draw a set of successively
inscribed spheres of appropriate radii for the scales $0 \ldots N$
chosen to cut off functions with coefficients below a predetermined
tolerance, and we keep in the basis only those functions of a given
scale which fall within the corresponding sphere (filled circles in
the figure).  We refer to this process as {\em
restriction}\index{Semicardinality!restriction}.  This process may be
carried out adaptively if desired\cite{tymczak}.

Such restriction maintains a description equivalent, within the
predetermined tolerance, to the full basis $V_N$  and thus the
underlying {\em uniform} basis on the finest scale.  In contrast to
finite element approaches, the basis functions of a multiresolution
analysis constructed this way do not move with the atoms.  The only
effect of the motion of the atoms is to turn on or turn off basis
functions whose coefficients are below the selected tolerance.

Figure~\ref{fig:C} shows two sets of results whose comparison
demonstrates how well this approach compacts the representation of
electronic wave functions.  The triangular points in the plot display
the results of an essentially exact calculation within the local
density approximation, obtained by exploiting the spherical symmetry
of the atom.  The curves in the figure show that the results obtained
from a full, three dimensional calculation using a restricted
multiresolution analysis are nearly indistinguishable from the exact
results.  Yet, rather than requiring a coefficient for each point of
the effective grid of $16$~million points, the results in the figure
come from a restricted basis with six levels of refinement giving a
total of only 2,500 functions, compacting the representation by over
three orders of magnitude!  

\section{Expression of physical couplings} \label{sec:SemiC}

\subsection{Non-linear, local couplings}

\begin{figure}
(a) \parbox{0.45\textwidth}{\includegraphics[width=0.43\textwidth]{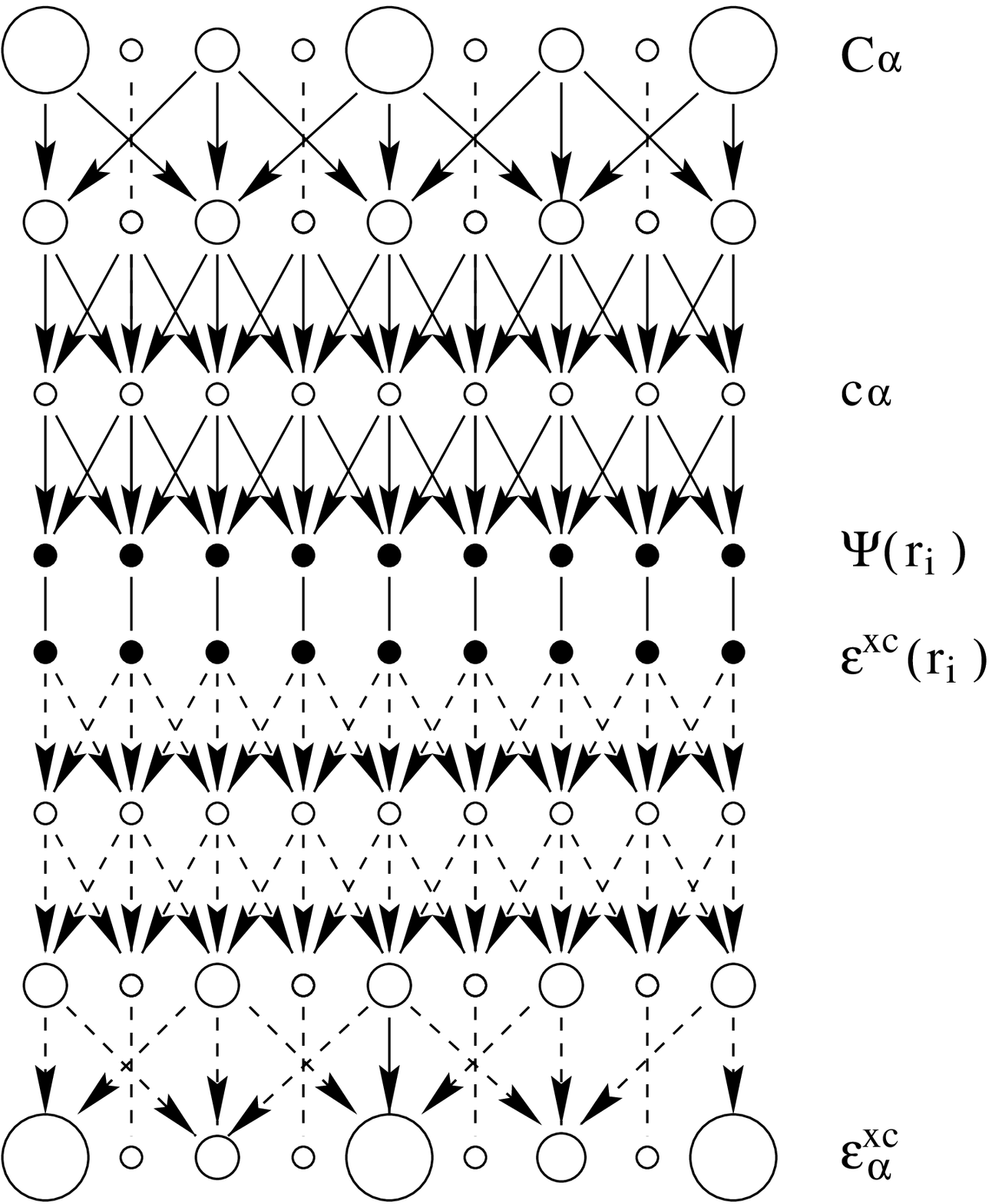}} (b) \parbox{0.45\textwidth}{\includegraphics[width=0.43\textwidth]{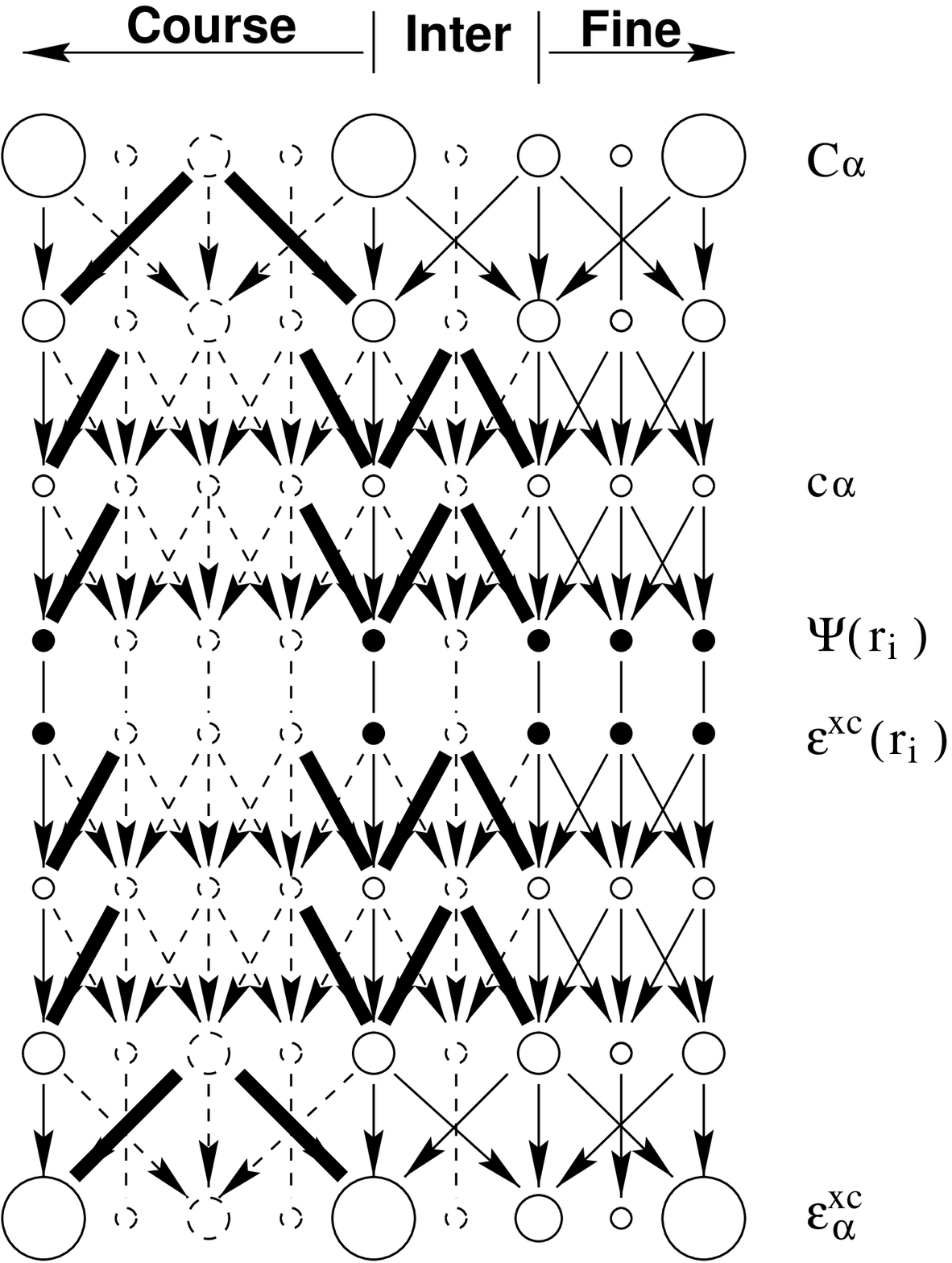}}
\caption{Calculation of the exchange-correlation field.  ({\bf a})
Unrestricted multiresolution analysis. ({\bf b}) Restricted multiresolution
analysis.  Coarse-scale expansion coefficients ({\it large
circles\,}), intermediate-scale coefficients ({\it mid-sized
circles\,}), fine-scale coefficients ({\it small circles\,}), function
values ({\it filled points\,}), multiply-accumulate operations ({\it
arrows\,})}
\label{fig:wt}
\end{figure}

Figure~\ref{fig:wt}a gives an explicit, diagrammatic representation of
the calculation of a non-linear, local coupling among physical fields.
Although the procedure is completely general, as an example we
consider the computation of the exchange-correlation energy per
particle $\epsilon^{\mbox{xc}}(\vec{r})$ in terms of the electronic
wave functions.  The calculation begins at the top of the figure with
the multiscale expansion coefficients $C_\alpha$ for the electronic
wave functions and ends at the bottom with the corresponding
multiscale coefficients for the exchange-correlation field.  The
circles represent values during the course of the calculation, and the
arrows represent mathematical operations which first multiply the
value at the base of the arrow by a coefficient specific to the arrow
and then accumulate the result onto the value at the head of the
arrow.  In the first and final rows, the large circles represent the
expansion coefficients for the coarse scaling functions in the
multiresolution analysis, the intermediate circles are those for the
detail functions and the small circles for the finer detail functions.

The calculation proceeds in three phases.  The first, {\em forward
phase} (first through fourth rows of the figure) determines the values
of the fields on a set of points in space $\vec{r}_i$ by first finding
the equivalent expansion coefficients $c_\alpha$ for the underlying
basis of finest-scale scaling functions and then using these
coefficients to compute the real-space values.  The first stage of
this process (first through third rows) is the fast wavelet transform,
explained below in Sec.~\ref{sec:fwt}.  The second stage (third
through fourth rows) involves taking the single-scale expansion
coefficients $c_\alpha$ (small circles), multiplying by the values of
the basis function on the points $\vec{r}_i$ and accumulating the
result onto those points (filled circles), as represented by the
arrows.  In the second, {\em coupling} phase (fourth through fifth
rows), the known values at the points $\vec{r}_i$ are combined in the
purely local, non-linear fashion prescribed by the coupling.  In this
example, at each point $\vec{r}_i$ we take the square magnitude
$|\psi(\vec{r}_i)|^2$ of each wave function, add the results together
to compute the charge density $n(\vec{r}_i)$ and then evaluate the
non-linear function $\epsilon^{\mbox{xc}}(n(\vec{r}_i))$.  The figure
represents this independent, point-by-point operation as a set of
simple vertical lines.  Finally, once the non-linear field is known at
all points in space, for the third, {\em inverse phase} of the
calculation (fifth through final row), one inverts the first stage of
the procedure and determines the multiscale expansion coefficients for
the field in terms of its real-space values by inverting each stage of
the forward phase in reverse sequence.  Inspection of each of the
forward stages reveals that it is a convolution, the inverse of which
is another convolution, which may be implemented in the same form but
with the arrows potentially carrying different coefficients.  (For a
more detailed discussion, we refer the reader to \cite{newrmp}.)

\subsection{Fast wavelet transform} \label{sec:fwt}

The upper three rows of Figure~\ref{fig:wt}a illustrate the {\em fast
wavelet transform}\index{Wavelets!fast wavelet transform}, which
computes the single-scale expansion coefficients $c_\alpha$ of a
function from its multiscale expansion coefficients $C_\alpha$.  The
basic strategy is for each stage of the calculation to give an
equivalent expansion of the original function but in terms of a
multiresolution analysis involving one fewer scale, proceeding in this
way until the single-scale representation is reached.

The upper two rows of the figure illustrate the first such stage.  As
Eqs.~(\ref{eqn:tsr},\ref{eqn:df}) prescribe, each coarse scaling
function and detail function of the original multiresolution analysis
(large and intermediate circles of the first row) may be expanded
exactly in terms of the intermediate scaling functions (larger circles
of the second row).  Thus, the arrows which multiply the values
carried by the two sets of larger circles on the first row by the
corresponding two-scale $c_n$ or detail $d_n$ coefficient and
accumulate the result onto the scaling functions of the next scale
compute the required equivalent expansion of the original function in
a multiresolution analysis of one fewer scale.  The coefficients of
the finer detail functions play no role in this process, and so are
simply copied to the second row as indicated by dashed lines in the
figure.  Each stage proceeds exactly in this manner.

The number of floating point operations in this transform varies
strictly linearly with the number of points in the grid, and thus
scales superiorly even to the fast Fourier transform.  Typically, the
fast wavelet transform requires only approximately 60 floating point
operations per grid point.  This efficiency, however, is quickly lost
when using a restricted multiresolution analysis.

\subsection{Impact of restriction}

Figure~\ref{fig:wt}b illustrates the impact of restricting the
multiresolution analysis on the calculation of local, non-linear
couplings.  The basis functions indicated by dashed circles are to be
removed from the basis, providing a coarse resolution on the left, an
intermediate level of resolution near the center, and a fine
resolution on the right.  The arrows show the same flow of information
as before, but now are drawn to convey the impact of the restriction
on the progress of the calculation.  Under the restriction, there are
three distinct types of information flow.  The dashed arrows carry
information onto values which have been restricted from the basis, and
so we may ignore them in the calculation.  The thin, solid arrows
carry information from functions in our basis to other functions in
our basis and represent calculations which we must perform.  Finally,
the thick, solid arrows are problematic and carry information from
functions restricted from the basis onto function maintained therein
and thus have the potential to corrupt the final results of the
calculation.

In practice, the thick arrows in the forward phase of the transform
(upper portion of the figure) do not corrupt the calculation
significantly because the expansion coefficients at their base were
determined to be small when the restriction for the basis was chosen.
The difficulty is in lower portions of the figure, where the arrows
carry not expansion coefficients, but {\em real-space values} of the
fields, which need not be small.  Tracing back the flow from these
arrows, it is evident that to properly compute the final coarse
expansion coefficients ultimately requires computing the the
non-linear coupling on all points of the finest grid.

In terms of computational expense, this sacrifices much of the benefit
of the multiresolution analysis.  Returning to the example of the
calculation of the carbon atom, although the wavelet transforms
require only about 60 operations per grid point and only 2,500 basis
functions were required in the basis, the need to evaluate fields on
the finest scale of 16~million points requires the expenditure of
400,000 floating point operations per basis function.  A
pseudopotential calculation (at 44~rydberg cutoff) in the same cell
also requires approximately 2,500 basis functions, but, through the
use of the fast Fourier transform, requires only approximately 400
floating point operations per basis function.  Without a major
advance, the computational expense of representing the physics of the
core in a multiresolution analysis thus would require such a
computational cost as to relegate such calculations to a very special
niche where systematic results are needed very badly.

Our strategy to make the wavelet transform competitive in {\em
restricted} multiresolution analyses is to find multiresolution
analyses which eliminate all information flow along the thick arrows
in Figure~\ref{fig:wt}b.  We could then compute the results of the
transform by simply ignoring the values associated with the dashed
circles in the figure so that the transforms may be computed with only
approximately 60 floating point operations per {\em basis function},
less effort than even the fast Fourier transform.  Profoundly, without
the thick diagonal arrows carrying information from finer scale
functions onto neighboring coarser functions, the results obtained in
a restricted multiresolution analysis of finite, variable resolution
then would be identical to {\em exact} results obtained by beginning
with full knowledge of the physical fields, transforming onto a grid
of arbitrarily fine resolution, computing the interaction, and
transforming the result back!  We refer to this property as {\em
exactness}.\index{Semicardinality!exactness}

\subsection{Exactness through semicardinality}

\begin{figure}
(a) \includegraphics[width=0.45\textwidth]{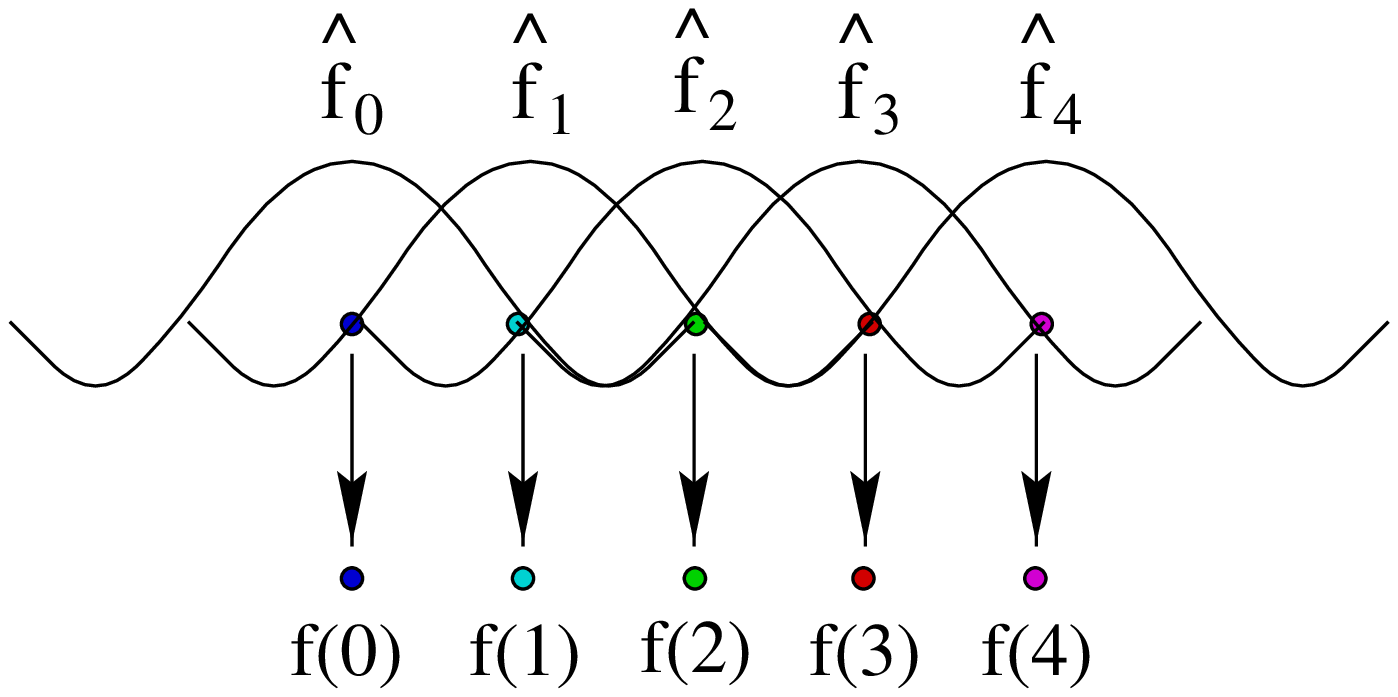}\\
(b) \includegraphics[width=0.45\textwidth]{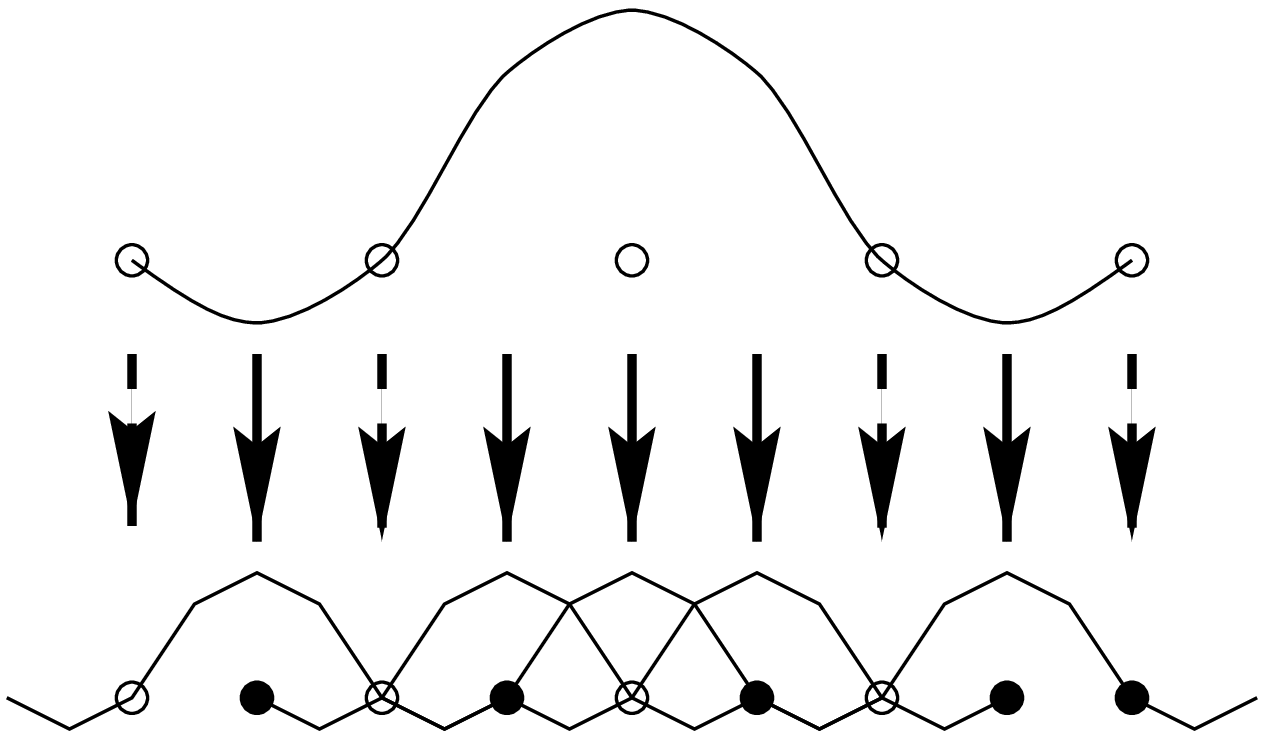}
(c) \includegraphics[width=0.25\textwidth]{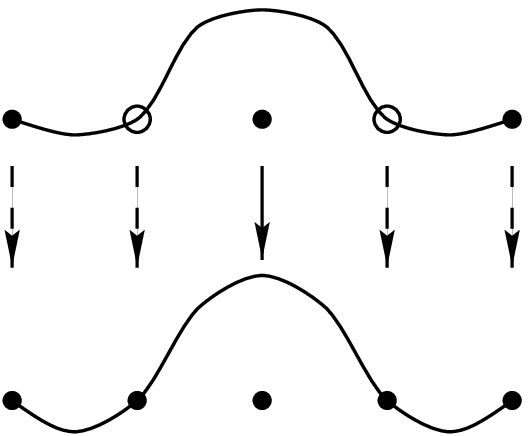}
\caption{Semicardinal multiresolution analysis.  ({\bf a}) Underlying uniform
basis of scaling functions.  ({\bf b}) Two-scale decomposition of scaling
functions, (c) decomposition of detail functions.  Zero coefficients
({\it dashed arrows\,})}
\label{fig:semicard}
\end{figure}

We now show a multiresolution analysis which gives rise to such exact
results for non-linear, local couplings.  For the scaling functions,
we draw inspiration from finite-element theory and consider {\em
cardinal\,}\index{Semicardinality!cardinality} functions, those with
zero value on all grid points except for one, where their value is
normalized to unity.  As Figure~\ref{fig:semicard}a illustrates, the
nodal properties of a single-scale basis of cardinal functions ensure
that the expansion coefficients of a function and its real-space
values on the grid are equal.  For the two-scale relation
(Figure~\ref{fig:semicard}b), this implies that the two-scale
coefficients are simply the values of the scaling function sampled on
a grid of half-spacing so that the two-scale coefficients never carry
information from a coarse function onto points associated with other
coarse functions, an important condition in obtaining exactness.

For the detail functions, we impose the condition that they be zero on
the points of the finer scale whose details they carry
(Figure~\ref{fig:semicard}c).  One cannot restrict the detail
functions to be appear cardinal on the points of the finer scales
because then they would become $\delta$-like functions.  The
coarser-looking, ``one-way'' form of the cardinality of the basis we
construct in this way we refer to as {\em
semicardinality}\index{Semicardinality}.  As
Figure~~\ref{fig:semicard}c illustrates, semicardinality implies that
all detail coefficients $d_n$ but one are zero, so that the detail
functions are simply finer-scale scaling functions, a condition not
exhibited in traditional wavelet bases.

\begin{figure}
\parbox{0.48\textwidth}{
(a) \includegraphics[width=0.47\textwidth]{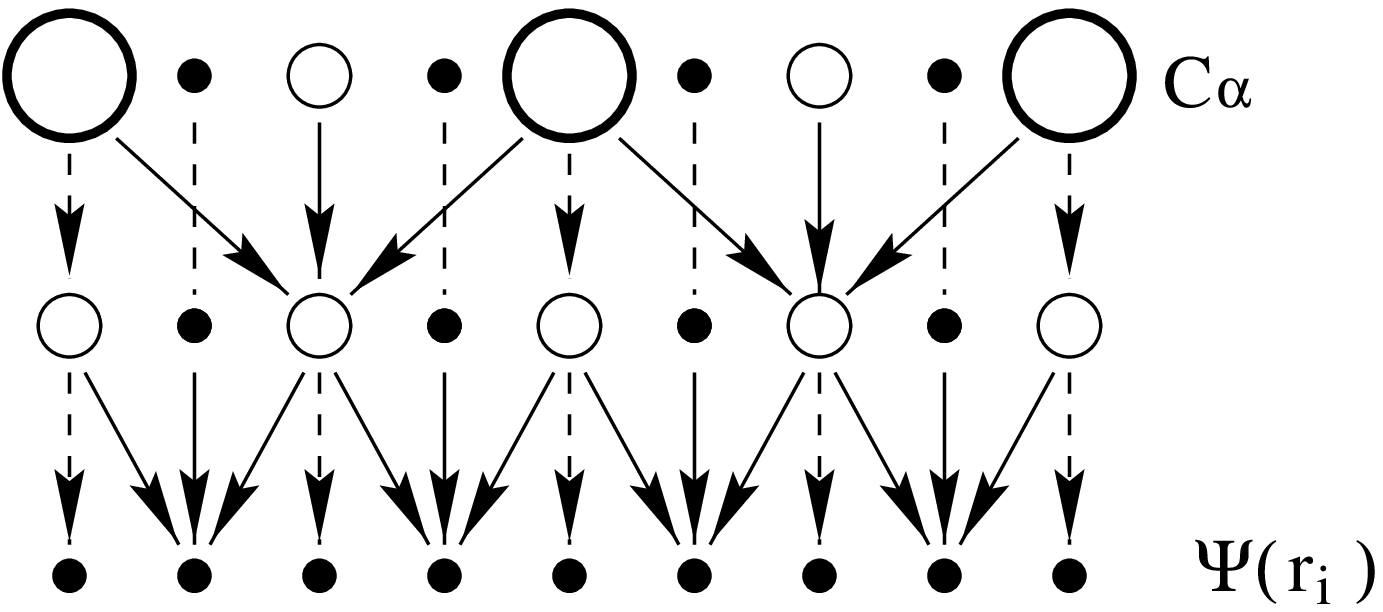}\\
\ \\
(b) \includegraphics[width=0.47\textwidth]{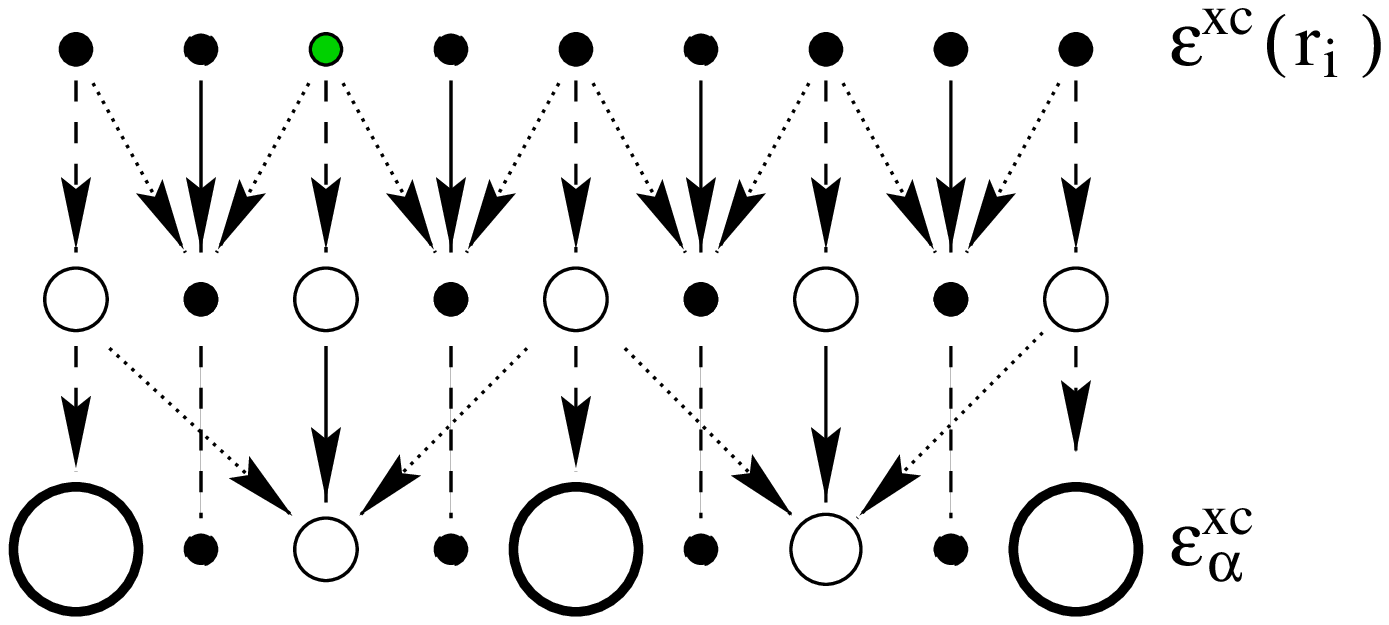}
}
\parbox{0.48\textwidth}{
(c) \includegraphics[width=0.47\textwidth]{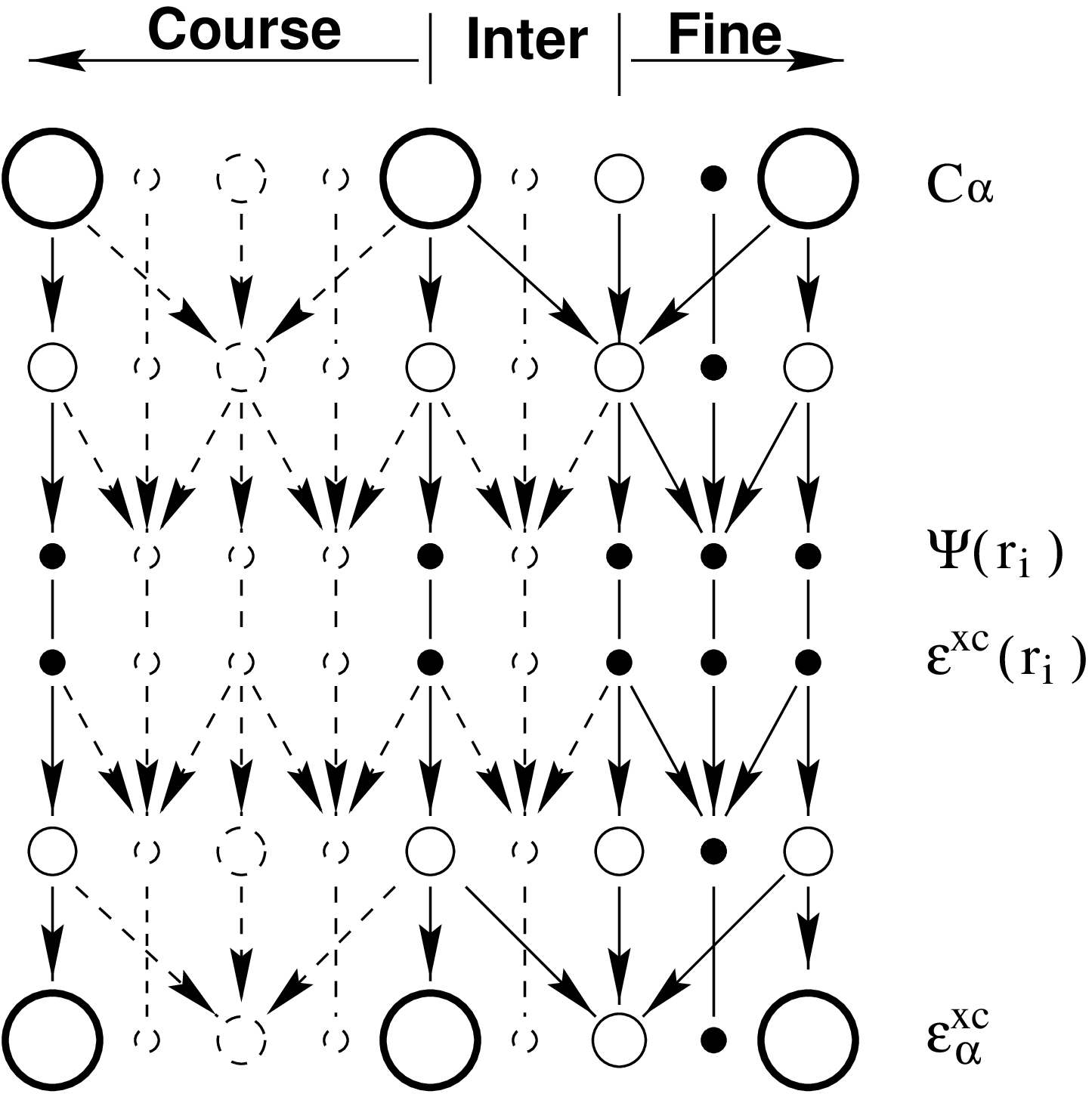}
}
\caption{Calculation of non-linear, local interactions in semicardinal
multiresolution analyses. ({\bf a}) Forward transform.  ({\bf b})
Inverse transform. ({\bf c}) Complete calculation in restricted basis}
\label{fig:semicardtrans}
\end{figure}

An immediate benefit of the cardinality of the scaling functions in a
semicardinal basis is that once the forward wavelet transform is
complete (Figure~\ref{fig:semicardtrans}a), the expansion coefficients
on the finest scale are just the values of the function on the finest
grid and the forward phase of the calculation is complete.  The nodal
conditions in the two-scale and detail coefficient sequences make
calculation of the inverse phase of the calculation simple as well.
Because at each stage of the forward transform
(Figure~\ref{fig:semicardtrans}a) the information flowing out from the
scaling functions carries vertically downward without change, in the
stages of the inverse transform (Figure~\ref{fig:semicardtrans}a) the
final values for the coarser functions may simply be copied again
vertically without change.  The information flowing out from the
detail functions in the forward transform flows vertically, but is met
by information flowing diagonally from the scaling functions.
Therefore, in the inverse transform, the final values for the detail
functions may be copied vertically but met with diagonal information
flow of the opposite sign.  Note that the pattern of information flow
is identical for the corresponding stages of the two transforms.  The
only change is reversal of the signs of the diagonal arrows. 

Figure~\ref{fig:semicardtrans}c shows the computation of a non-linear,
local interaction in a restricted {\em semicardinal} multiresolution
analyses.  The expression of these interactions involves no corrupting
information flow and is now {\em
exact\,}\index{Semicardinality!exactness} and may be computed more
efficiently than with the fast Fourier transform!  Similar results
hold in semicardinal bases for the economical expression of couplings
through differential operators.  One also may compute these
interactions {\em exactly} while working with only those functions
that survive a restriction\cite{newrmp}.

\section{Efficient solutions} \label{sec:solutions}

\subsection{Poisson's equation}

To solve Poisson's equation\index{Solution of equations!Poisson}, we
use the variational principle of minimizing the electrostatic
(Hartree) energy.  When the potential $\phi(\vec{r})$ is expanded in a
basis, the result is a linear system of equations involving the
Galerkin representation for the Laplacian operator.  In this section,
we focus on the convergence rate of conjugate gradient techniques in
solving the resulting equation.

As our example, we compute the electrostatic potential of the carbon
atom as a full three-dimensional problem.  The basis we employ
consists of 12 levels of refinement, each consisting a cubic array of
approximately $24^3$ detail functions.  (We use cubic refinement
regions as a matter of computational expediency.)  The total number of
functions in the basis is approximately $200,000$, and the basis
represents a range in length scale and thus condition number of the
Laplacian operator in excess of $2^{12}$ and $2^{24} \approx 16 \times
10^6$, respectively.  Under these conditions, one would expect
conjugate gradients to require tens of thousands of applications of
the operator to solve the equation.

\begin{figure}
\includegraphics[width=0.55\textwidth]{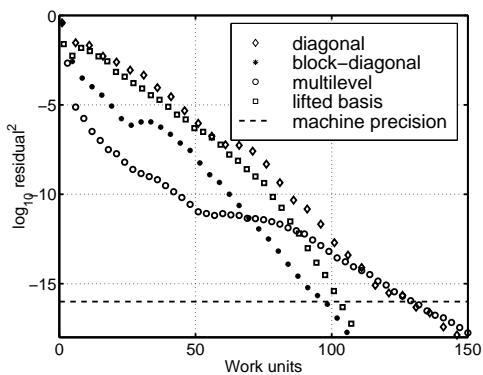}
\caption{Solution of Poisson's equation.  Convergence of square magnitude of
residual vector with various preconditioners in semicardinal and
lifted multiresolution bases}
\label{fig:poissonconv}
\end{figure}

Figure~\ref{fig:poissonconv} shows the convergence of conjugate
gradient methods for solving Poisson's equation with different
preconditioners: ``diagonal'' multiplies the residual by the inverses
of the diagonal matrix elements of the Laplacian, ``block-diagonal''
uses Fourier techniques to invert the diagonal sub-blocks of the
Laplacian matrix which connect functions of the same scale with each
other, and ``multi-level'' uses special techniques to invert the upper
and lower triangular blocks of the Laplacian matrix connecting
functions to functions of only higher or lower scales, respectively.
These three sets of calculations were carried out using a semicardinal
bases of third order.  The data labeled ``lift'' are results for
simple diagonal preconditioning of the Galerkin representation of the
Laplacian in a second type of basis, a basis of third-order {\em
lifted wavelets}\index{Wavelets!lifted wavelets} of the Sweldens
construction\cite{sweldens:96}.  The vertical axis of the figure
displays the square-magnitude of the residual vector, which varies
linearly with the error in the computation of the electrostatic energy
because the calculation is variational.  To accurately reflect the
varying complexity of the different preconditioners, the horizontal
axis indicates net computational time in ``work units'' (WU) defined
to be the computational effort of applying the Laplacian operator in
the semicardinal basis.  For the simple preconditioners, the
horizontal axis closely corresponds to number of iterations.

The upshot of the figure is that in these multiresolution bases,
conjugate gradients determines the electrostatic energy to within
machine precision in only about 100 iterations, three orders of
magnitude better than expected!  (The first reported calculations
which uses this basis with the diagonal preconditioner appear in
\cite{mgras}.)  These very high convergence rates, $\sim
0.15$~digits/WU, reflect the fact that the semicardinal basis
reorganizes the structure of the operator in such a way to enhance its
diagonal structure.  The most successful approach is the
block-diagonal preconditioner.  Whereas the multi-level preconditioner
gives a dramatic improvement in the number of iterations required to
solve the equation, these benefits are outweighed by its computational
cost.

In regards to calculation with lifted wavelets, it has been suggested
recently that the non-zero integral of the detail functions of a
semicardinal basis would create a bottleneck in the solution of
Poisson's equation and that, instead, one should use a lifted
interpolet basis in which the low order moments of the detail
functions vanish\cite{goedecker}.  Figure~\ref{fig:poissonconv}b ({\it
squares\,}) shows that although there is a modest improvement in the
convergence at the final stages of the calculation, there is no great
advantage in the solution of Poisson's equation from the properties of
the lifted basis.

\begin{figure}
(a) \parbox{0.45\textwidth}{\includegraphics[width=0.42\textwidth]{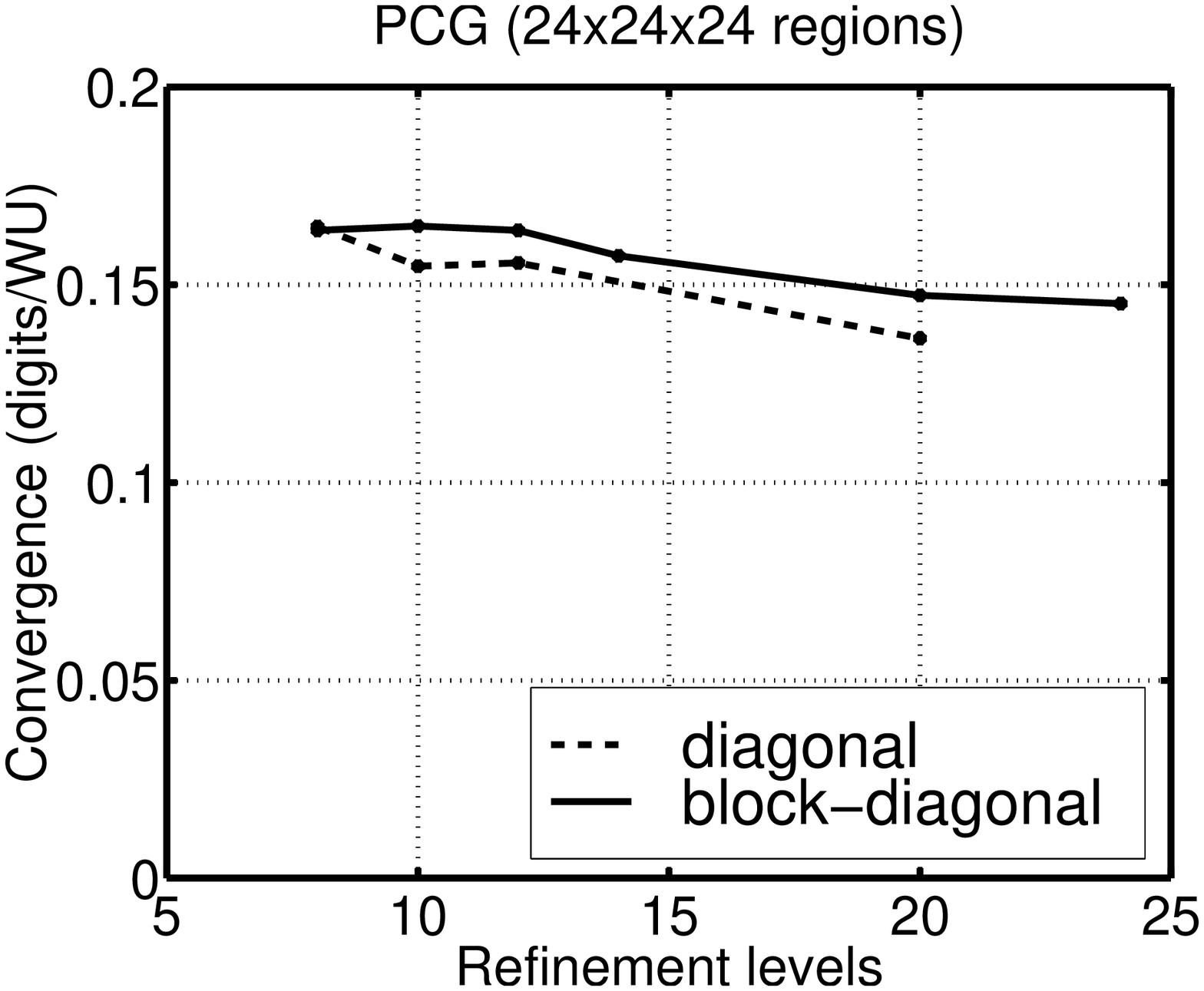}}
(b) \parbox{0.45\textwidth}{\includegraphics[width=0.42\textwidth]{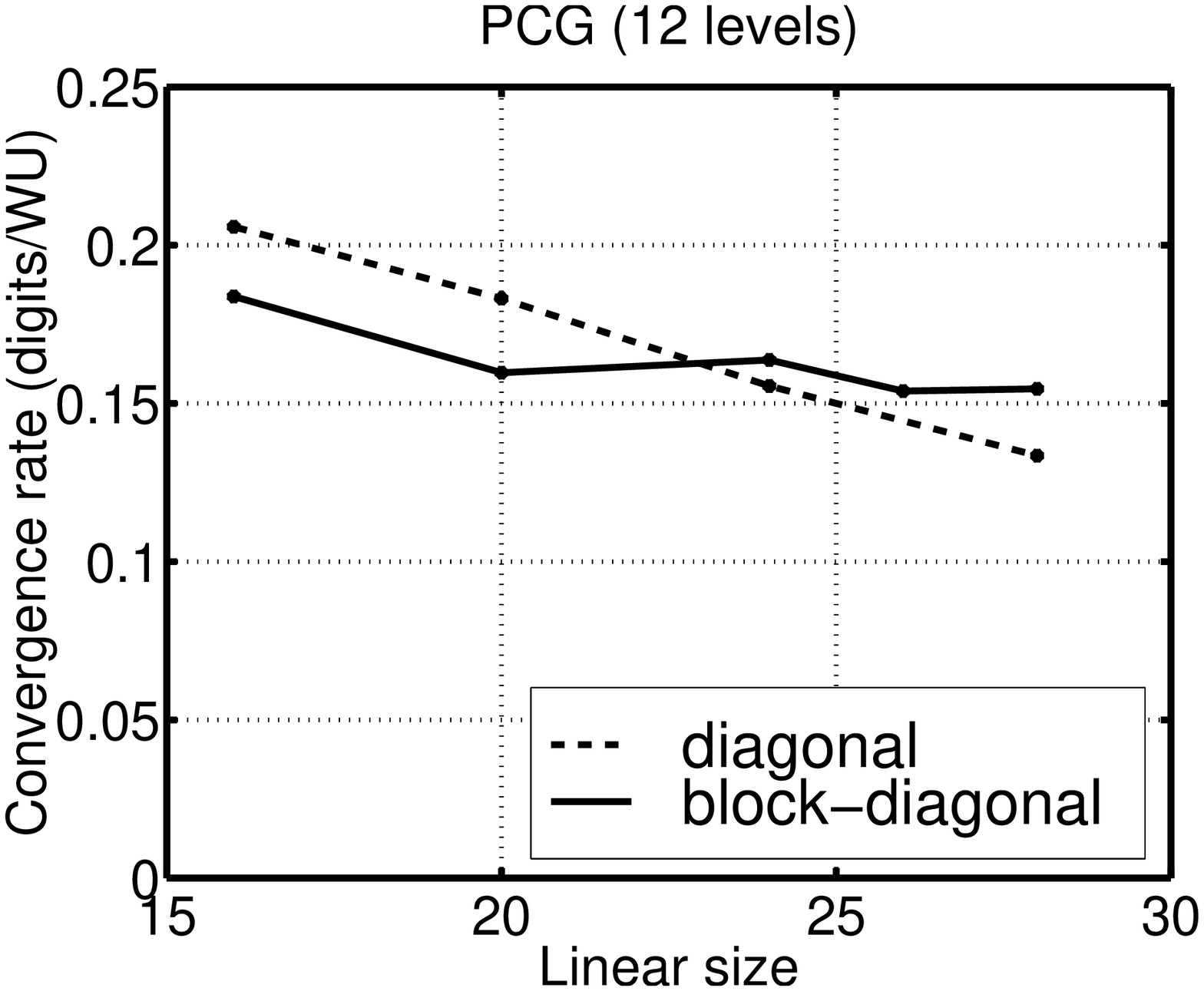}}
\caption{Behavior of convergence rate in solution of
Poisson's equation in semicardinal bases. ({\bf a}) Variation with number of
refinement regions, each of size $24^3$. ({\bf b}) Variation with spatial
extent of refinement regions}
\label{fig:poissonconvII}
\end{figure}

Finally, we consider the impact of problem size on the convergence
rate of the electrostatic energy as we increase the level of
resolution and the spatial extent of the refinement regions
(Figures~\ref{fig:poissonconvII}a~and~\ref{fig:poissonconvII}b,
respectively).  Figure~\ref{fig:poissonconvII}a shows that going as
far as 24 levels of refinement, the convergence rate of the diagonal
and block-diagonal preconditioners does not degrade appreciably
despite the condition number of the Laplacian operator now reaching
$2^{48} \sim 280$~trillion.  This finest level places 1,000 grid
points across the extent of the carbon nucleus, more than sufficient
to resolve the individual quarks within the nucleons.
Figure~\ref{fig:poissonconvII}b explores the behavior of convergence
rate as we increase the linear extent of the refinement regions up to
about two times the size expected to be relevant in the calculation of
electronic structure.  Although the diagonal preconditioner degrades
somewhat over this range, the block-diagonal conditioner maintains a
nearly constant convergence rate of 0.16~digits/WU for the
electrostatic energy.  The stability of these convergence rates
strongly suggests that, regardless of problem size, a fixed number of
approximately 100 iterations are required to solve Poisson's equation
to a given tolerance.

\subsection{Kohn-Sham equations} \label{sec:Lagrsolve}

As we saw in the preceding section, the solution of Poisson's equation
in semicardinal bases is surprisingly efficient, but still requires on
the order of one hundred iterations.  This is particularly burdensome
in the calculation of electronic structure, where in traditional
formulations the Poisson equation must be solved at each step in the
minimization of the Kohn-Sham energy functional.  We propose instead
to search directly for the stationary point of the Lagrangian
functional (\ref{eqn:saddle}).\index{Solution of equations!Kohn-Sham}
From the signs of the semi-local couplings, it is clear that the
stationary point of the Lagrangian functional is a minimum with
respect to the electronic wave function fields $\{\psi_i(\vec{r})\}$
and a maximum with respect to the electrostatic potential field
$\phi(\vec{r})$.  Although we are searching for a saddle point, our
proposed approach is analogous to iterative conjugate gradient
minimization but with a few key differences.  Each iteration begins at
a state point $\left(\{\psi_i(\vec{r})\},\phi(\vec{r})\right)$, but,
rather than searching along the downhill-gradient direction
$\left(-\nabla_{\psi} \cL,-\nabla_{\phi} \cL\right)$, we search along
the direction $\vec{d}\equiv \left(-\nabla_{\psi} \cL,\nabla_{\phi}
\cL\right)$, which is downhill for the electrons and uphill for the
electrostatic potential.  Next, rather than searching to the point
where the functional becomes stationary along the line defined by
$\vec{d}$, we search for the point where the next search direction
computed in this same way becomes perpendicular to $\vec{d}$.

\begin{figure}
\includegraphics[width=0.55\textwidth]{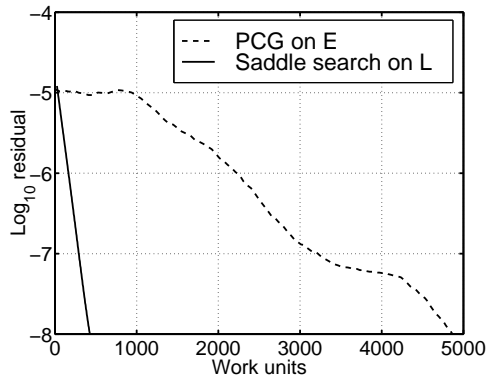}
\caption{Comparison of energy and Lagrangian functional solutions to
Kohn-Sham equations of density functional theory}
\label{fig:Lagr}
\end{figure}

For the beryllium atom, Figure~\ref{fig:Lagr} compares the convergence
of this new approach with the traditional approach of applying
preconditioned conjugate gradients to minimize the energy functional.
The horizontal axis measures computational in work units representing
the cost of the application of the Laplacian operator.  The figure
shows that the computational savings from not having to solve
Poisson's equation at each iteration makes the Lagrangian approach a
full order of magnitude more efficient than the traditional approach
based upon minimization of the energy functional.


\clearpage
\addcontentsline{toc}{section}{Index}
\flushbottom
\printindex

\end{document}